\documentclass{article}
\usepackage{epsfig}

\date{\today}

\usepackage{amsmath}

\newcommand{\la}{\lambda}
\newcommand{\ka}{\kappa}
\newcommand{\al}{\alpha}

\newcommand{\ta}{\theta}
\newcommand{\f}{\phi}

\newcommand{\ee}{\end{equation}}
\newcommand{\eea}{\end{eqnarray}}
\newcommand{\be}{\begin{equation}}
\newcommand{\bea}{\begin{eqnarray}}

\newcommand{\pa}{\partial}

\newcommand{\vep}{\varepsilon}

\newcommand{\re}[1]{(\ref{#1})}

\begin{document}
\begin{center}

{\Large Spinning $U(1)$ gauged Skyrmions}
\vspace{0.6cm}
\\
 Eugen Radu$^{\dagger}$
and  D. H. Tchrakian$^{\dagger \star}$
\\
$^{\dagger}$
{\small \it Department of
Mathematical Physics, National University of Ireland Maynooth,}
\\
$^{\star}${\small \it School of Theoretical Physics -- DIAS, 10 Burlington
Road, Dublin 4, Ireland }
\end{center}

\begin{abstract}
We construct  axially symmetric solutions of 
U(1) gauged Skyrme model.
Possessing a nonvanishing magnetic moment, these solitons have also a
nonzero angular momentum proportional to the electric charge.
\end{abstract}
\medskip

\noindent{\textbf{Introduction.--~}}
Many nonlinear classical field theories on flat spacetime backgrounds admit 
soliton solutions. These nonsingular solutions
describe particle-like, localised configurations with finite energy.
There has been some interest in recent years in the 
issue of globally regular spinning soliton solutions.
However, to the best of our knowledge, to date no {\it stationary} and
{\it spinning} solitons were found. 
 (We describe single lumps with angular momentum
as spinning, and reserve rotating for more general (gravitating-) solutions, including multilumps.) 
Notably, it is known that that
finite energy solutions of the Yang-Mills-Higgs (YMH) system
with a nonvanishing magnetic charge have zero angular
momentum~\cite{VanderBij:2001nm, vanderBij:2002sq} \footnote{Also the axially
symmetric spinning Einstein--Yang-Mills sphalerons,
although predicted perturbatively \cite{Brodbeck:1997ek}, 
are unlikely to exist \cite{VanderBij:2001nm, Kleihaus:2002ee}, but
these are in anycase not topologically stable.}. 
Moreover, as found in \cite{Volkov:2003ew}, none of the known gauge field
solitons with gauge  group $SU(2)$ ($e.g.$ dyons, sphalerons, vortices) admit
spinning generalizations within the stationary,
axially symmetric, one-soliton sector. 
 
To date two types of spinning solitons have been found in the literature,
{\it a}) $Q$-balls solitons in a complex scalar field theory with a
non-renormalizable self-interaction \cite{Volkov:2002aj}, which are
nontopological solitons so their stability is not guaranteed by a topological
charge, and {\it b}) the electrically charged dipole
monopole--antimonopole pair \cite{Paturyan:2004ps} of the YMH system 
with vanishing topological charge, which is not topologically stable
even in the limit of vanishing angular momentum.

It is our purpose here to construct a soliton which
has intrinsic angular momentum and presents a
  {\it topologically stable limit}\,\footnote{An axially 
symmetric, spinning soliton of the ungauged Skyrme model, similarly presenting
a {\it topologically stable limit},
has been recently constructed in \cite{Battye:2005nx}. However, this is a
$Q$-ball type of solution featuring time-dependent fields.}. Our definition
for a 'soliton presenting a topologically stable limit' is, a finite
energy spinning lump which is topologically stable
in the limit of vanishing angular momentum.
This configuration corresponds to axially symmetric, electrically charged
solutions of the $U(1)$ gauged Skyrme model.

Concerning the question of the existence of any given topologically stable
solution, this is quite an intricate matter that deserves a brief description.
To start with, there must be a valid topological lower bound on the energy,
which may or may not be saturated, and for the Skyrmion it is not. Then there
is the question whether any given field configuration
(the solution) does minimise the energy? For the Skyrme model, this is a
difficult problem for two reasons: {\it a}) because the sigma model fields are
constrained, and {\it b}) because in addition to the quadratic kinetic term
there is also a quartic kinetic term. Thus for the $1$-Skyrmion, the existence
proof is given by \cite{Esteban} and, \cite{LinandYang}, while for axially
symmetric case, to the best of our knowledge, 
there is no rigorous existence proof. So axially symmetric
Skyrmions and their magnetically gauged counterparts are supported only
numerically.

In addition, when a nonvanishing electric field is present, as it is in the
present work, the functional misnimised is not the positive definite energy
but the indefinite action. The proof of existence of such solutions, namely
that for YMH dyons, is given by \cite{SchechterandWeder}, but again it is too hard
to adapt this proof for the gauged (and ungauged) Skyrme model.  Thus the
existence of the $U(1)$ gauged axially symmetric solutions
of the present paper, and those of \cite{Battye:2005nx}, are supported only
numerically. 

\medskip
\noindent{\textbf{The model.--~}}
The Skyrme model has been proposed a long time ago \cite{Skyrme:1961vq}
as an effective theory for nucleons in the large $N$ limit 
of QCD at low energy~\cite{Witten:1983tx,Witten:1983tw,Adkins:1983ya},
the baryon number being identified with 
the topological charge.
The classical as well as the quantum properties are 
in relatively good agreeement with the observed features of small nuclei.
The $U(1)$ gauged Skyrme model
was originally proposed by Callan and Witten to study the
decay of the nucleons in the vecinity of a monopole \cite{Callan:1983nx}.
Axially symmetric solutions of this model were constructed
previously in \cite{Piette:1997ny}, but the emphasis 
there was on the static properties of
nucleons and not the calculation of its classical spin.

We define our model
in terms of the $O(4)$ sigma model field $\f^a=(\f^{\al},\f^A)$, $\al=1,2$;
$A=3,4$, satisfying the constraint $|\f^{\al}|^2+|\f^A|^2=1$, the Lagrangean
of the Maxwell gauged Skyrme model is
(up to an overall factor which we set equal to one)
\be
\label{lagu1}
{\cal L}=-\frac12|F_{\mu\nu}|^2+\frac12|D_{\mu}\f^a|^2
-\frac{\ka^2}{8}\,|D_{[\mu}\f^a\,D_{\nu]}\f^b|^2
\ee
in terms of the Maxwell field strength $F_{\mu\nu}$, and the covariant
derivatives defined by the gauging prescription 
\be
\label{u1pres}
D_{\mu}\f^{\al}=\pa_{\mu}\f^{\al}+A_{\mu}\,(\vep\f)^{\al}\quad,\quad
D_{\mu}\f^A=\pa_{\mu}\f^A\,.
\ee
The energy-momentum tensor which follows from (\ref{lagu1}) is
\bea
T_{\mu\nu}&=&\Bigg\{-2\left(F_{\mu\la}\,F_{\mu}^{\la}
-\frac14\,g_{\mu\nu}\,F_{\tau\la}\,F^{\tau\la}\right)+
\left(D_{\mu}\f^a\,D_{\nu}\f^a
-\frac12\,g_{\mu\nu}\,D_{\la}\f^a\,D^{\la}\f^a\right)\nonumber\\
&-&2\cdot\frac{\ka^2}{4}\left[\left(D_{[\mu}\f^a\,D_{\la]}\f^b\right)
\left(D_{[\nu}\f^a\,D^{\la]}\f_b\right)-\frac14\,g_{\mu\nu}
\left(D_{[\tau}\f^a\,D_{\la]}\f^a\right)
\left(D^{[\tau}\f_a\,D^{\la]}\f_b\right)\right]\Bigg\}\,.\label{stress-mn}
\eea
Here we note that the Skyrmion gauged with
the purely magnetic $U(1)$ field is a topologically stable soliton. This is
stated in terms of topological lower bound on the static energy density
functional of the purely magnetically gauged system, namely the $T_{tt}$
component of \re{stress-mn} with $A_t=0$,
\be
\label{enfinconv}
T_{tt}={\cal E}=|F_{ij}|^2+|D_i\f^a|^2
+\frac{\ka^2}{4}\,|D_{[i}\f^a\,D_{j]}\f^b|^2\,.
\ee
Defining the gauge invariant topological charge density as
\bea
\varrho&=&\frac{1}{4\pi}\vep_{ijk}\vep^{abcd}\,\,
D_i\f^a\,D_j\f^b\,D_k\f^c\,\f^d+\frac{3}{8\pi}\vep_{ijk}\,F_{ij}\,
\vep^{AB}\,\f^B\,\pa_k\f^A
\label{ch2s}
\\
&=&\frac{1}{4\pi}\vep_{ijk}\vep^{abcd}\,\,
\pa_i\f^a\,\pa_j\f^b\,\pa_k\f^c\,\f^d-\frac{3}{4\pi}\vep_{ijk}\pa_k\left(
\,A_i\,\vep^{AB}\,\pa_j\f^A\,\f^B\right)\label{ch1s}
\eea
the gauge invariance of $\varrho$ is manifest from \re{ch2s}, while it is
easily checked that the finite energy conditions lead to the vanishing of
the surface integral term in \re{ch1s}, as a result of which the topological
is simple the volume integral of the first term, namely the winding number $n$
or, Baryon charge.

As was shown in \cite{Piette:1997ny} in detail, the energy density functional
\re{enfinconv} is bounded from below by
\be
\label{lb}
{\cal E}\ge\frac{\ka}{\sqrt{1+\frac94\ka}}\,\varrho\,.
\ee
\\
\noindent{\textbf{The ansatz.--~}}
In a cylindrical coordinate system,
we parametrise the axially symmetric Maxwell connection as
\be
\label{axu1}
A_t= b(\rho,z)\quad,\quad
A_{\al}=\frac{a(\rho,z)-n}{\rho}\,\ \vep_{\al\beta}\,\ \hat x_{\beta}
\quad,\quad A_z=0\,,
\ee
$a(\rho,z)$ and $b(\rho,z)$ corresponding to the electric and magnetic
potentials, with $n$ a positive integer - the winding number,
and the polar parametrisation of the chiral field in terms of the two
functions $f(\rho,z)$ and $g(\rho,z)$ as
\bea
\label{axf}
\f^{\al}=\sin f\sin g\ n^{\al},
~~~
\f^3=\sin f\cos g ,
~~~\f^4=\cos f,
\eea
where $\rho=\sqrt{|x_{\al}|^2}$, $\al=1,2$, and $z=x_3$. 
In the following we
will find it convenient instead to work with spherical coordinates $(r,\ta)$,
i.e. $\rho=r\sin\ta$ and $z=r\cos\ta$.
After replacing this ansatz in (\ref{lagu1}), one finds the reduced lagrangeean
\begin{eqnarray}
\nonumber
L&=& r^2 \sin \theta \Bigg\{
\frac{2}{r^2\sin^2\ta}\left(a_{,r}^2+\frac{1}{r^2}a_{,\ta}^2\right)-
2 \left(b_{,r}^2+\frac{1}{r^2}b_{,\ta}^2\right) 
\\
\label{2dfconstrrtjz}
&+&
\left[\left(f_{,r}^2+\frac{1}{r^2}f_{,\ta}^2\right)
+\left(g_{,r}^2+\frac{1}{r^2}g_{,\ta}^2\right)\sin^2f+
\frac{a^2-r^2\,b^2\sin^2\ta}{r^2\sin^2\ta}\sin^2f\sin^2g\right]+
\\
\nonumber
&+&
\ka^2\sin^2f\left(\frac{1}{r^2}\left(f_{,r}g_{,\ta}-f_{,\ta}g_{,r}\right)^2+
\frac{a^2-r^2\,b^2\sin^2\ta}{r^2\sin^2\ta}
\left[\left(f_{,r}^2+\frac{1}{r^2}f_{,\ta}^2\right)
+\left(g_{,r}^2+\frac{1}{r^2}g_{,\ta}^2\right)\sin^2f\right]\sin^2g
\right) \Bigg\}.
\end{eqnarray}
The Euler-Lagrange equations arising from the variations of
this Lagrangean have been integrated by imposing the following boundary
conditions, which respect finite mass-energy and finite energy density
conditions as
well as regularity and symmetry requirements. We impose 
\begin{eqnarray}
\label{infty}
f|_{r=\infty}=0,~~g_{,r}|_{r=\infty}=0,~~
a|_{r=\infty}=n,~~b|_{r=\infty}=V,
\end{eqnarray}
at infinity, and 
\be
\label{orig}
f|_{r=0}=\pi,~~ g_{,r}|_{r=0}=0,~~a|_{r=0}=n,~~  b_{,r}|_{r=0}=0,
\ee
at the origin.
For solutions with parity reflection symmetry (the case considered in
this paper), the boundary conditions along the $z$-axis are
\begin{eqnarray}
\label{tapi2MS}
  f_{,\ta}|_{\theta=0}=g|_{\theta=0}=0,
~~  a_{,\ta}|_{\theta=0}=  b_{,\ta}|_{\theta=0}=0, 
\end{eqnarray}
and agree with the boundary conditions on the $\rho$-axis, 
except for $g(r,\theta=\pi/2)=\pi/2$.

 It may appear from the boundary conditions \re{infty}-\re{tapi2MS}
that the natural condition $a|_{\ta=0,\pi}=n$ is not imposed. This is not done
since its imposition in addition to \re{infty}-\re{tapi2MS} would be an
an overdetermination. We have nontheless checked that $a=n$ is satisfied
on the $z$-axis by the numerical solutions. 

The constant $V$ appearing in (\ref{infty}) corresponds to the magnitude of
the electric potential at infinity and has a direct physical relevance.
In the pure Maxwell theory, one can set set $V=0$
(or any other value) without any loss of generality. In the U(1) gauged Skyrme
model, however, such a gauge transformation would render the whole
configuration time-dependent.

Integration over all space of the energy density 
${\cal E}$ yields the total mass-energy,
$E = \int T_{tt} \sqrt{-g} d^{3}x.$
The  total angular momentum is given by
$J =\int T_{\varphi t}\sqrt{-g} d^{3}x,$
where
\begin{eqnarray}
T_{t\f}=  2\left(a_{,r}\,b_{,r}+\frac{a_{,\ta}\,b_{,\ta}}{r^2}\right)
+a\,b\,\,\sin^2f\sin^2g\left(1+\ka^2\left[
\left(f_{,r}^2+\frac{f_{,\ta}^2}{r^2}\right)+
\left(g_{,r}^2+\frac{g_{,\ta}^2}{r^2}\right)\sin^2f\right]\right).
\end{eqnarray}
However, by using the field equations, the volume integral of the above
quantity can be converted into a surface integral at infinity
in terms of Maxwell potentials
\begin{eqnarray}
J=4\pi \lim_{r \rightarrow \infty} \int_{0}^{\pi}d\theta \sin \theta~r^2 a~b_r.
\end{eqnarray}
The field equations imply the  asymptotic behaviour of
the  electric potential 
$b\sim V-Q/(2r)+O(1/r^2),$ the parameter
$Q$ corresponding to the electric charge of the solutions.
Therefore the following relation holds
\begin{eqnarray}
\label{JQ}
J=4 \pi nQ,
\end{eqnarray}
which resembles the case of a monopole-antimonopole configuration in a 
YMH theory  \cite{Paturyan:2004ps}.
Note that the solutions discussed here 
possess also a magnetic dipole moment  \cite{Piette:1997ny} 
which can be read from the asymptotics of the $U(1)$ 
magnetic potential, $A_{\varphi}\sim \mu \sin\theta/r^2$.
 

\newpage
\setlength{\unitlength}{1cm}
\begin{picture}(6,8)
\centering
\put(2,0){\epsfig{file=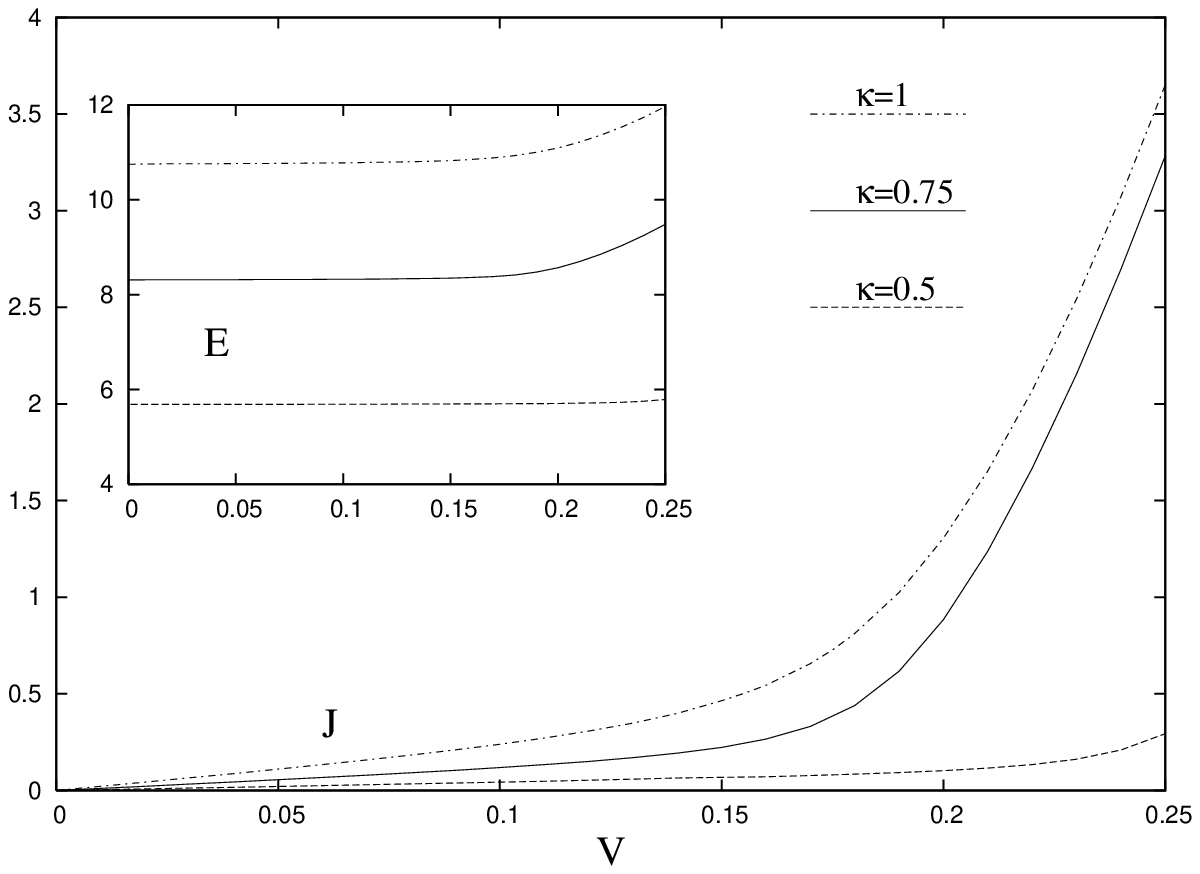,width=12cm}}
\end{picture}
\begin{center}
Figure 1a.
\end{center}
\begin{picture}(20,8.5)
\centering
\put(2.7,0){\epsfig{file=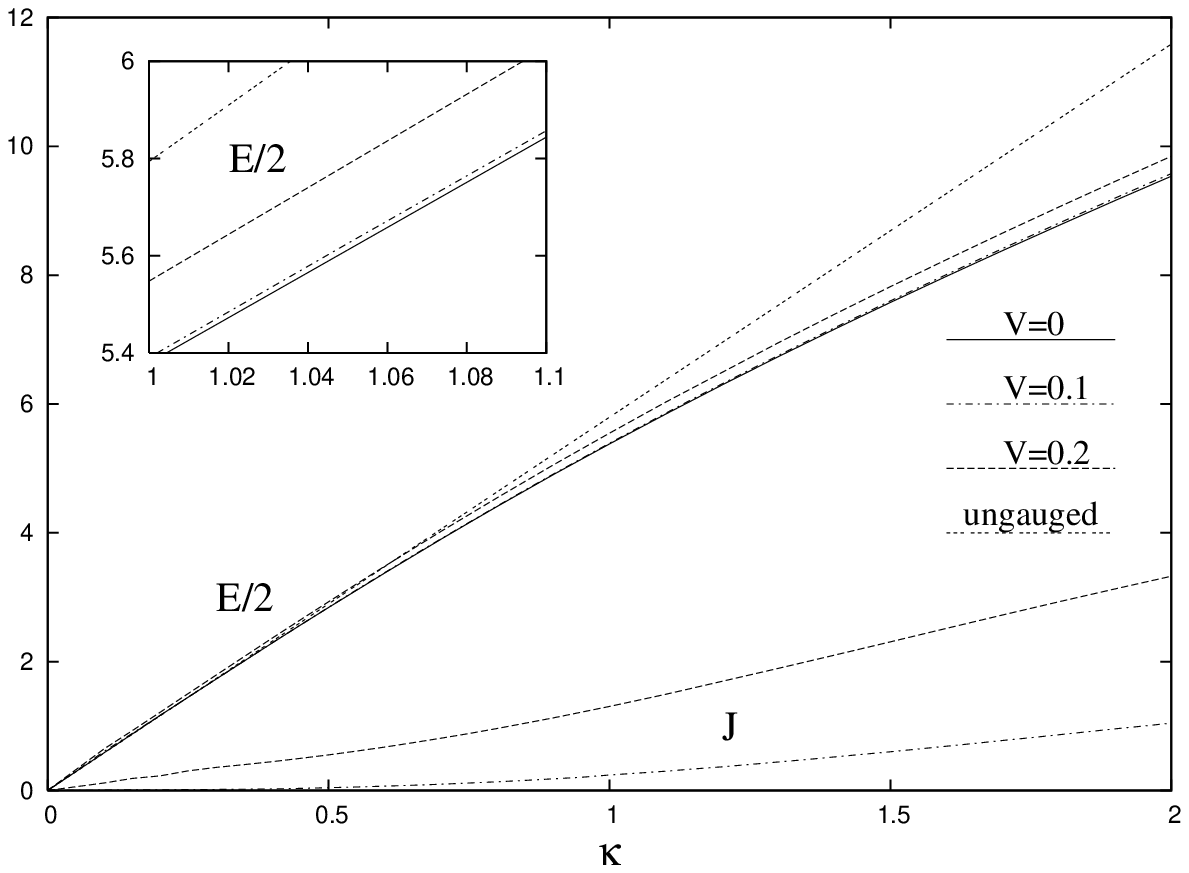,width=12cm}}
\end{picture}
\begin{center}
Figure 1b.
\end{center}
{\small {\bf Figure 1.} The energy $E$ and the angular momentum $J$ 
of U(1) gauged Skyrmion 
are shown as a function on the parameter $V$ (Figure 1a) 
and the parameter $\kappa$ (Figure 1b) for a baryon number $n=1$.}
\\
\\
\noindent{\textbf{Numerical solutions.--~}}
Subject to the above boundary conditions,
we solve  numerically the set of four Maxwell-Skyrme equations.
The numerical calculations are performed by using the program
CADSOL \cite{FIDISOL}, based on the iterative Newton-Raphson method.
As initial guess in the iteration procedure, we use the spherically symmetric
regular solutions of the pure Skyrme model.
The typical relative error  is estimated to be 
lower than $10^{-3}$.

\newpage
\setlength{\unitlength}{1cm}
\begin{picture}(6,6)
\centering
\put(2,0){\epsfig{file=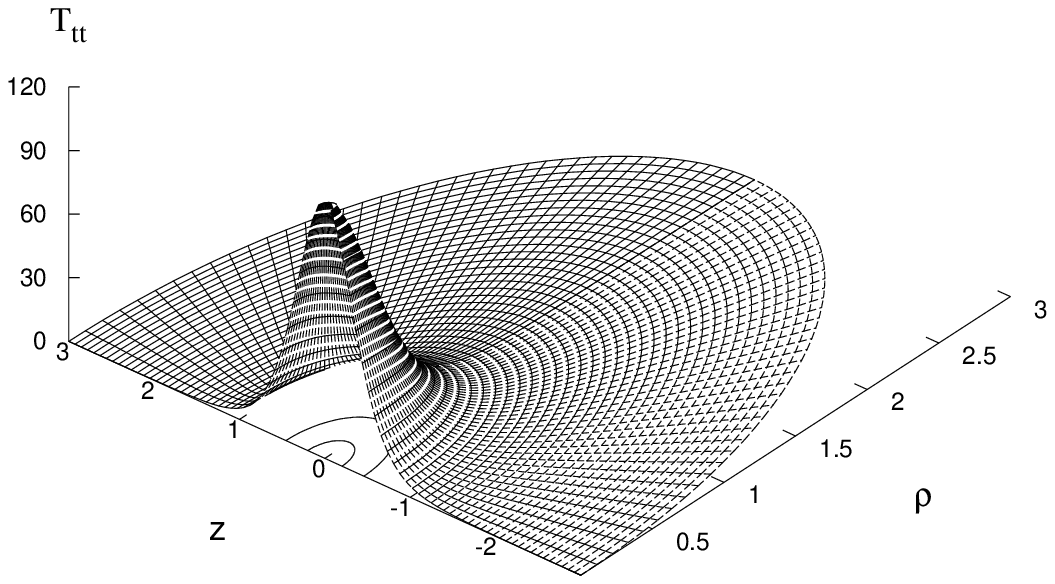,width=11.5cm}}
\end{picture}
\begin{center}
Figure 2a.
\end{center}
\begin{picture}(18,7.)
\centering
\put(2.7,0){\epsfig{file=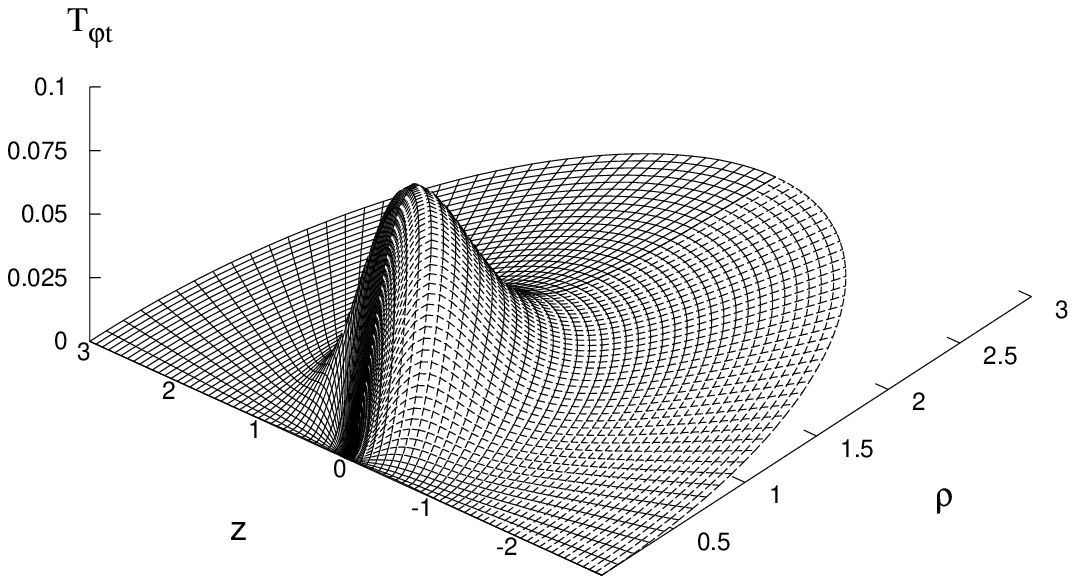,width=11.5cm}}
\end{picture}
\begin{center}
Figure 2b.
\end{center}
{\small {\bf Figure 2.} The components $T_{\varphi t}$ and $T_{tt}$ of the
energy momentum tensor are shown for
 a typical $n=1$ solution, with $\kappa=0.72,~V=0.067$.}
\\
\\
For a given Baryon number, 
the solutions depend on two continuos parameters, the values $V$
of the electric potential at infinity and the 
Skyrme coupling constant $\kappa$.
Here we consider solutions in the one baryon sector only, although
similar results have been found for $n>1$.
The solutions with $V=0$ have $b=0$ and correspond 
to static dipoles discussed in \cite{Piette:1997ny}.
A nonvanishing $V$ leads to rotating regular configurations,
with nontrivial functions
$f,~g,~a$ and $b$.
Rotating solutions appear to exist for any value of $\kappa$.
As we increase $V$ from zero while keeping 
$\kappa$ fixed, a branch of solutions forms.
Along this branch, the total energy and the angular momentum  
increase continuously with $V$.
The ration $J/E$ increases also, but remains always 
smaller than one.
At the same time, the numerical errors start to increase 
and we obtain large values for
both $E$ and $J$, and for some $V_{max}$ the numerical
iterations fail to converge. 
An accurate value of $V_{max}$ is rather difficult to obtain, 
especially for large values of $\kappa$.
Alternatively, we may keep fixed the magnitude of the electric
potential at infinity and vary the parameter $\kappa$. 

In Figure 1 we present the properties of typical branches of solutions. In
Figure 1a, the angular momentum and the energy are parametrised by $V$ for
several fixed value of $\kappa$, while in Figure 1b these quantities are
parametrised with $\ka$ for several fixed values of $V$, including $V=0$
corresponding to the non-spinning soliton.   The energy bound in the
purely magnetically gauged case with $V=0$ is not saturated, as is the
case also for the ungauged Skyrmion. We expect likewise that this numerically
constructed solution is topologically stable, but cannot estimate the energy
excess above the lower bound analytically.  

One can see from Figure 1b that, for a given value of $\kappa$,
the energy of the spinning soliton is always smaller than
the energy of the ungauged Skyrmion, but is larger than the 
energy of the corresponding non-spinning static gauged solution.   The
latter is gauged only with the magnetic field and minimises the energy
functional, while the spinning system gauged with both the magnetic and the
electric fields minimises the non-positive definite Lagrangian density,
and the additional electric field does not feature in the topological
lower bound. As a result, the spinning, electrically charged, solutions
have higher energies than the static ones. The situation here is identical with
that of the Julia-Zee dyon, in this respect. 

In Figure 2a we plot the energy density ${\cal E}=T_{tt}$, and in Figure 2b
the angular momentum density $T_{\varphi  t}$
of a typical $n=1$ solution as a function of the
coordinates $\rho, z$, for $\kappa=0.72,~V=0.067$.
We notice that the energy density $\epsilon=T_{tt}$ does not
exhibit any distinctly localised individual components, a surface
of constant energy density being
topologically a sphere. However, this is a deformed sphere such that the
profiles of ${\cal E}=T_{tt}(r,\ta)$ versus $r$ for each value of $\ta$
are distinct and non overlapping. It presents a peak on the symmetry axis,
and the density profiles decrease monotonically with $r$.

Also, the electrically charged $U(1)$ gauged Skyrmion rotates as a
single object and the $T_{\varphi  t}$--component of the energy momentum
tensor associated with rotation presents a maximum in the $z=0$ plane and
no local extrema (see Figure 2b).

\medskip
\noindent{\textbf{Conclusions.--~}}
We have presented here the
first example of  spinning solution residing
in the one-soliton sector of the theory which has a
topologically stable limit.   
These solutions of the $U(1)$-gauged Skyrme model
carry mass, angular momentum, electric charge and a magnetic
dipole momentum. The electric charge is induced by rotation and
equals the total angular momentum.

Similar qualitative results have been found by adding to the Lagrangean
(\ref{lagu1}), a self-interaction
potential of the $O(4)$ scalar field representing the pion mass.
Nonzero pion masses lead to larger values for the energy and angular momentum.
 
Also, we have found that similar to the ungauged case, the spinning Skyrmions 
admit gravitating generalisations, which are currently under study.
These solutions satisfy also the generic relation (\ref{JQ}).
\\
\\
\noindent
{\bf\large Acknowledgements}
\\
 We are indepted to the referee for
instructive comments, and to Yisong Yang for his generous advice.
We are also grateful to Burkhard Kleihaus for useful remarks on a first version of this paper.
\\ 
This work is carried out
in the framework of Enterprise--Ireland Basic Science Research Project
SC/2003/390.

\end{document}